\documentclass[prl,aps,preprintnumbers,twocolumn, acknowledgement]{revtex4}
\usepackage{amsmath,amssymb,bm}
\usepackage{graphicx}
\usepackage{color}

\renewcommand{\v}[1]{{\bf #1}}

\newcommand{\s}{{\sigma}}

\def\eqa{\begin{eqnarray}}
\def\eea{\end{eqnarray}}
\newcommand{\eq}{\begin{equation}}
\newcommand{\ee}{\end{equation}}
\newcommand{\nn}{\nonumber\\}
\newcommand{\Eq}[1]{Eq.~(\ref{#1})}
\newcommand{\<}{\langle}
\renewcommand{\>}{\rangle}

\newcommand{\ua}{\uparrow}
\newcommand{\da}{\downarrow}
\newcommand{\ra}{\rightarrow}

\newcommand{\del}{\delta}
\newcommand{\Del}{\Delta}
\newcommand{\eps}{\epsilon}
\newcommand{\veps}{\varepsilon}

\newcommand{\ka}{\kappa}
\newcommand{\la}{\lambda}

\newcommand{\si}{\sigma}

\newcommand{\cH}{ {\cal H} }

\begin{document}
\title{Tuning topological orders by a conical magnetic field in the Kitaev model}
\author{Ming-Hong Jiang$^{1}$}
\author{Shuang Liang$^{1}$}
\author{Wei Chen$^{1,2}$} \email{chenweiphy@nju.edu.cn}
\author{Yang Qi$^{3,4,2}$}
\author{Jian-Xin Li$^{1,2}$}
\author{Qiang-Hua Wang$^{1,2}$} \email{qhwang@nju.edu.cn}
\affiliation{$^{1}$National Laboratory of Solid State Microstructures and School of Physics, Nanjing University, Nanjing, China}
\affiliation{$^2$Collaborative Innovation Center of Advanced Microstructures, Nanjing University, Nanjing, China}
\affiliation{$^3$Center for Field Theory and Particle Physics, Department of Physics, Fudan University, Shanghai 200433, China}
\affiliation{$^4$State Key Laboratory of Surface Physics, Fudan University, Shanghai 200433, China}

%\date{\today}
\begin{abstract}
We show that a conical magnetic field $\v H=(1,1,1)H$ can be used to tune the topological order and hence anyon excitations of the $\mathrm{Z_2}$ quantum spin liquid in the isotropic antiferromagnetic Kitaev model. A novel topological order, featured with Chern number $C=4$ and Abelian anyon excitations, is induced in a narrow range of intermediate fields $H_{c1}\leq H\leq H_{c2}$. On the other hand, the $C=1$ Ising-topological order with non-Abelian anyon excitations, is previously known to be present at small fields, and interestingly, is found here to survive up to $H_{c1}$, and revive above $H_{c2}$, until the system becomes trivial above a higher field $H_{c3}$. The results are obtained by devoloping and applying a $\mathrm{Z_2}$ mean field theory, that works at zero as well as finite fields, and the associated variational quantum Monte Carlo. 
\end{abstract}

\maketitle  

{\em Introduction}: The Kitaev model is one of prototype models that support unambiguously quantum spin liquid states~\cite{Kitaev2006}, the states of matter under wide and intense investigations~\cite{Anderson1973, Anderson1987, Lee2008, Balents2010, Zhou2017}. The model is exactly solvable in the absence of applied magnetic field~\cite{Kitaev2006}. The value of such a toy model is it may describe several universal families of topological states of matter. The stability of the spin liquids under perturbations beyond the exactly solvable model, such as the magnetic field, is of both theoretical and experimental interest~\cite{Zhu2017, Liang2018, Gohlke2018, Trebst2018, Jiang2018, He2018, Trivedi2018, Gordon2019, Wen2017, Ran2017, Wang2017, Yu2018, Ponomaryov2017}. For example, on top of the gapless B-phase of the model at zero field, perturbation theory for a small conical magnetic field results in an effective topological Ising model~\cite{Kitaev2006,Nayak2011}, or effective chiral $p_x+ip_y$-wave superconductor,\cite{Read2000} featured with non-Abelian anyon excitations, which may have potential applications in topological quantum computing~\cite{Kitaev2006}.

Of intense recent interest is to investigate the stability of the Ising-topological state beyond the perturbative limit~\cite{Zhu2017,  Gohlke2018, Trebst2018, Jiang2018, He2018, Trivedi2018}. This is a highly nontrivial issue. On one hand, the Ising-topological state requires the magnetic field to break the time-reversal symmetry~\cite{Kitaev2006}. On the other hand, a sufficiently large magnetic field can suppress quantum fluctuations in favor of trivial polarized state. The questions to ask are: 1) Where does the Ising-topological spin liquid become unstable, and 2) Between the two limits of weak and strong magnetic fields, is there a new state of spin liquid? Several groups applied exact diagonalization (ED) for small clusters, and density matrix renormalization group (DMRG) for quasi-one-dimensional (1D) cylinders~\cite{Zhu2017,  Gohlke2018, Trebst2018, Jiang2018, He2018, Trivedi2018}. 
All of these works suggest a gapless U(1) spin liquid phase at intermediate magnetic fields, under the assumption that the two-dimensional (2D) spinon dispersion can be folded to 1D. However, the limited size accessible in ED and DMRG  makes it arguably hard to judge the intrinsic properties of the system in the thermodynamic limit. In fact, it is indicated in Ref.\cite{Zhu2017} that the large entanglement entropy found in DMRG implies long-range correlations that might be hard to capture by small-radius cylinders. 
It is therefore highly desirable to have an effective theory that can access the thermodynamic limit. Mean field theory (MFT) in terms of partons, or slave particles, is of this type~\cite{Wen2002, Wen2004}. For the Kitaev model under concern, an SU(2) MF solution was proposed at zero field~\cite{Nayak2011}. The extension to finite magnetic fields is plagued by the confining local-moment solution, hence only analytic perturbation theory for small fields is reported.~\cite{Kitaev2006,Nayak2011}

\begin{figure}
	\includegraphics[width=0.9\columnwidth]{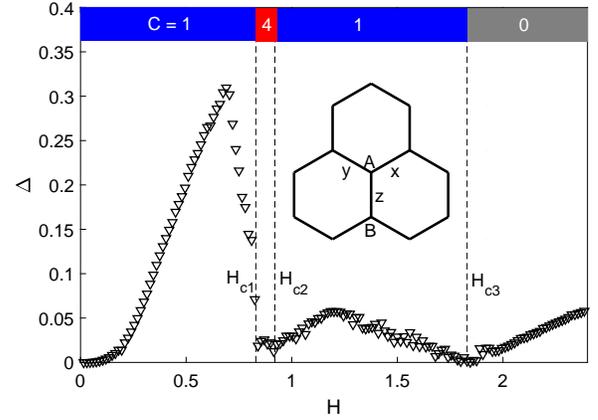} %{gap_vmc_honey.png}
	\caption{Topological order, characterized by the Chern number in the color bar, versus the applied magnetic field. Also shown is the spinon excitation gap (symbols), which closes at $H_{c1,c2,c3}$ (vertical dashed lines). The results are obtained from VMC-optimized effective Hamiltonian. The inset illustrates the Kitaev-honeycomb lattice. See the text for more details.}\label{fig:gap}
\end{figure}
  
Here we develop an efficient Z$_2$ MFT aplicable to all cases of the applied field, which turns out to be asymptotically exact and agrees with the associated variational Monte Carlo (VMC) excellently at finite fields for the Kitaev model. For a conical magnetic field $\v H=(1,1,1)H$ and from the VMC-optimized effective Hamiltonian, we find topological phase transitions with the Chern number changing from $C=1$ to $4$, $1$, and $0$, successively at $H_{c1,c2,c3}$, see Fig.\ref{fig:gap}. Therefore, the field induces a new topological order with $C=4$, in addition to the $C=1$ Ising-topological state. The latter is known from perturbation theory at small fields, and is found here to survive up to $H_{c1}$ and revive above $H_{c2}$, until the system becomes trivial eventually above $H_{c3}$.  

{\em Model and Method}: The Kitaev model under a magnetic field $\v H=(H_x,H_y,H_z)$ is described by the Hamiltonian
\eqa \cH = &&\sum_{a=x,y,z}\sum_{\<\v r\v r'\>\in a}J_a \hat{\si}_a(\v r) \hat{\si}_a(\v r')-\sum_{\v r a}H_a \hat{\si}_a(\v r). \label{eq:kitaev}\eea
Here $\hat{\si}_a(\v r)\equiv ic_a(\v r) c_0(\v r)$ is twice of the spin-1/2 operator $S_a(\v r)$ at position $\v r$ of the honeycomb lattice and along direction $a$. We also use $a$ to label the three sets of bonds $\<\v r\v r'\>$ radiating from the A-sublattice sites, see the inset of Fig.\ref{fig:gap}. The Majorana fermions, $c_{\mu}(\v r)$, with $\mu=0,x,y,z$, are normalized as $\{c_\mu(\v r),c_\nu(\v r')\}=2\del_{\mu\nu}\del_{\v r\v r'}$. To eliminate the redundancy in the Fock space, we need to require
\eqa Q_a(\v r) \equiv -i c_a(\v r) c_0(\v r) -\frac{i}{2} \sum_{bc}\eps_{abc}c_b(\v r) c_c(\v r)=0,\label{eq:charge} \eea
where $\eps$ is the $3\times 3$ completely anti-symmetric tensor. For reasons to be clearer shortly, we call $\v Q(\v r)$ the charge vector at $\v r$. The above charge neutrality constraint is equivalent to $D(\v r)\equiv c_0(\v r) c_x(\v r) c_y(\v r) c_z(\v r) =-1$ Kitaev originally proposed, but turns out to be much more convenient and more revealing as we shall show. 

The MF Hamiltonian can be written as
\eqa \cH_{MF}=&&-\sum_a\sum_{\<\v r\v r'\>\in a} \sum_{\mu,\nu=0,a}J_a K^a_{\mu\nu}(\v r,\v r') ic_{\bar\mu}(\v r) c_{\bar\nu}(\v r') \nn
&&-\sum_{\v r a} [ B_a(\v r) \hat{\si}_a(\v r) + \la_a(\v r)Q_a(\v r)] +\cdots.\label{eq:mmft}\eea
Here $K^a_{\mu\nu}(\v r,\v r')=i\<c_\mu(\v r) c_\nu(\v r')\>(2\del_{\mu\nu}-1)$, $\bar{\mu}=0$ (or $a$) if $\mu=a$ (or $0$), and similarly for $\bar{\nu}$ and $\nu$. The effective field $B_a(\v r)=H_a -J_a m_a(\v r')$ for $\v r'$ the neighbor of $\v r$ on bond $a$, with $m_a(\v r)=\<\hat{\si}_a(\v r)\>$. The local Lagrangian multiplier $\la_a(\v r)$ is used to enforce local charge neutrality on average. Finally, the dotted part is the irrelevant MF constants.

We can re-express the Majorana fermions in terms of complex fermions $a_{\ua,\da}$, dropping the argument $\v r$ wherever applicable: $c_x = a_\ua + a_\ua^\dag$, $c_y= i a_\ua - ia_\ua^\dag$, $c_z = -a_\da - a_\da^\dag$, and $c_0 = -i a_\da + i a_\da^\dag$. Then we have $Q_a = 2\psi^\dag\tau_a\psi$, where $\psi^\dag=(a_\ua^\dag,a_\da^\dag,a_\da,-a_\ua)$ is the twisted Nambu spinor, and $\tau_a$ is the Pauli matrix in the particle-hole basis. Up to a global factor, $Q_a$ measures exactly the deviation (in three ways) from the neutral single-occupancy of the $a_\si$-fermions, hence is indeed a charge operator. It can be shown that $[Q_a, \cH]$ is linear in $Q_{b\neq a}$ in the Majorana representation, hence \Eq{eq:kitaev} lacks local charge-SU(2) gauge symmetry, {\em but it is still a faithful representation of the original spin model in the invariant $\v Q(\v r)=0$ subspace}. The remaining local gauge symmetry is $Z_2$, $c_\mu(\v r)\ra s(\v r) c_\mu(\v r)$, where $s(\v r)=\pm 1$. We refer to the ensuring MFT stated above as the Z$_2$ MFT. Using the above mapping, we may rewrite Eq.\ref{eq:mmft} in the form 
\eqa \cH_{MF}&&= \frac{1}{2}\sum_{a\mu\nu}\sum_{\<\v r\v r'\>\in a}[(\psi^\dag(\v r) h_{\mu\nu}(\v r,\v r')\si_\mu\tau_\nu\psi(\v r') +{\rm H.c.}] \label{eq:cmft} \nn
&&+\frac{1}{2}\sum_{\v r a} \psi^\dag(\v r)[h_a(\v r)\si_a+\ka_a(\v r)\tau_a]\psi(\v r)+\cdots.\label{eq:su2mft}\eea
Here $\si_\mu$ is the Pauli matrix in the spin basis. The set of parameters, $\{h_{\mu\nu}(\v r,\v r'),h_a(\v r),\ka_a(\v r)\}$, is related to $\{K^a_{\mu\nu}(\v r),B_a(\v r),\la_a(\v r)\}$ in Eq.(\ref{eq:mmft}), see Ref.\cite{SM}.
Note the MF Hamiltonian describes spinon quasiparticles, while the energy of the system is more straightforwardly calculated by $E=\<\cH\>$ using Wick contractions. On the other hand, the $\si_0\tau_{x,y}$ components in $\cH_{MF}$ are singlet pairings, and the $\si_a\tau_{x,y}$ components are triplets. In particular, the local singlet pairing potentials $\ka_{x,y}(\v r)$ are needed in general to enforce the average local singlet pairing $\<Q_{x,y}(\v r)\>=0$. 

The complex-fermion representation is used to perform VMC. We borrow the structure of $\cH_{MF}$, but tune the parameters  to form a variational Hamiltonian $\cH_v$. The ground state $|G\>$ of $\cH_v$ can be written as
\eqa |G\>=[\sum_{\v r\si,\v r'\s'}a_{\si}^\dag(\v r) A(\v r\si,\v r'\si')a_{\si'}^\dag(\v r')]^{N/2}|0\>,\eea
where $N$ is the number of sites, and the kernel $A$ a skew matrix by fermion antisymmetry. The physical wavefunction for the quantum spins is $|G_s\> = \Pi_\v r (1-e^{i\pi n(\v r)})|G\>$, where $n(\v r)=\sum_\si a^\dag_\si(\v r) a_\si(\v r)$. The physical energy of the system is calculated as
$E = \<G_s|\cH |G_s\>/\<G_s|G_s\>$ by standard MC, and similarly for other observables. The energy is optimized by varying the parameters to obtain the best approximation to the ground state of the physical quantum spins. More technical details can be found in Ref.\cite{SM}.

\begin{figure}
	\includegraphics[width=0.9\columnwidth]{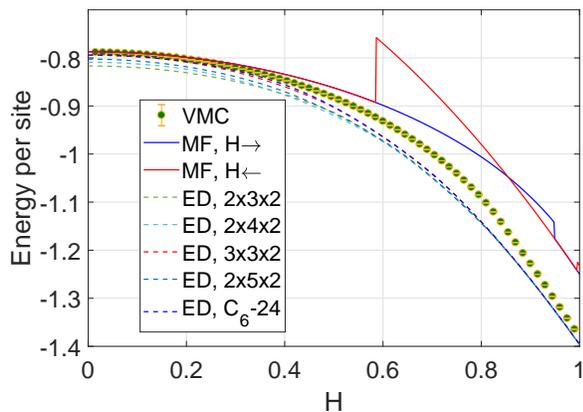} %.png
	\caption{The energy per site versus $H$ from MFT (solid lines), VMC (symbols) and ED of small clusters of various sizes (dashed lines). The field increases (decreases) on blue (red) solid lines in MFT.}\label{fig:energy}
\end{figure}

{\em Results and discussions}: The MFT and VMC described above can be applied in general cases, but from now on we will limit ourselves to the specific case: $J_a=J=1$ and $H_a=H$, or $\v H=(1,1,1)H$. The solid lines in Fig.\ref{fig:energy} shows the energy per site versus $H$ from MFT. The field increases (decreases) on the blue (red) solid line during the iterative MF calculations, using the previous result as the initial condition for the next field. There is a level crossing at $H=H_{c,MF}\sim 0.855J$, suggesting a first order phase transition at this field. Below $H_{c,MF}$, the uniform MF ground state follows the blue solid line, with the following structure of parameters in \Eq{eq:cmft}: $ h_a(\v r) = h_\si$, $\ka_a(\v r) = h_\tau$, and for $\<\v r\v r'\>\in a$ and $\v r$ on the A-sublattice, $h_{\mu\nu}(\v r,\v r') = i [t_0\del_{\mu 0}+t_{||}\del_{\mu a}+t_\perp (1-\del_{\mu 0})(1-\del_{\mu a})]\del_{\mu\nu}$. We denote the set of five parameters as
\eqa X=\{h_\si,h_\tau,t_0,t_{||},t_{\perp}\}.\label{eq:ansatz}\eea 
(In fact, in the uniform MF solution we find $t_{||}=-t_0$, but we will take them independent in VMC later to enlarge the optimization space, without violating the $C_3$ lattice symmetry.)

The MF state on the red solid line at $H>H_{c,MF}$ is a local moment state with $h_{\mu\nu}(\v r,\v r')=0$ and nonuniform $h_a(\v r)$ (and is highly degenerate due to the internal frustration). This is a purely classical state. Although it appears to have lower energy at $H>H_{c,MF}$ in the MF sense, it can not be the true ground state due to the absence of any quantum fluctuations.

We check the robustness of the MFT by VMC, by tuning the parameters in Eq.(\ref{eq:ansatz}). The symbols in Fig.\ref{fig:energy} show the optimized energy versus $H$. (The results are obtained in a $N=6\times 6\times 2$-site lattice, and we checked that the finite-size effect is insignificant.) Note the statistical error bar is well within the symbol size, and in particular, the error vanishes (independent of the number of samples used) as $H\ra 0$. In fact it turns out that there is nothing to optimize at $H=0$. The projected MF state is the exact ground state of the spin liquid at $H=0$. Because of this asymptotic exactness, the MFT deviates from VMC only mildly at finite fields. The relative error between MFT and VMC for $H<H_{c,MF}$ is below $10\%$. This is already remarkable, as compared to the situation for the Heissenberg model, where the (unrenormalized) MF energy would be just about a quarter of the VMC result.\cite{fuchun} At higher fields, we find the ansatz in Eq.(\ref{eq:ansatz}) continues to work in VMC, without running into any first-order transition. This is because by implementing the constraint exactly, VMC can capture quantum fluctuations beyond MFT. It is very reasonable that such fluctuations can drive the MF first-order transition into a crossover. 

While VMC is statistically exact, it is subject to the bias built in the ansatz Eq.(\ref{eq:ansatz}) for the trial ground state $|G\>$. For this reason, we further check the energetics by ED for small periodic clusters up to 24 sites. The results are presented as dashed lines in Fig.\ref{fig:energy}. With increasing lattice size, the ED energy approaches the VMC result very well at both low and high magnetic fields, particularly for the $3\times 3\times 2$-cluster and the $C_6$-symmetric 24-site cluster. Even in the intermediate field regime, the relative departure is at most $5\%$. Taken together, the VMC under the ansatz Eq.(\ref{eq:ansatz}) works to our satisfaction. More results from ED are available in Ref.\cite{SM}, which are inconsistent with the gapless U(1) spin liquid phase assumed to lie between two fixed critical fields.\cite{Zhu2017,  Gohlke2018, Trebst2018, Jiang2018, He2018, Trivedi2018}

\begin{figure}
 \includegraphics[width=0.75\columnwidth]{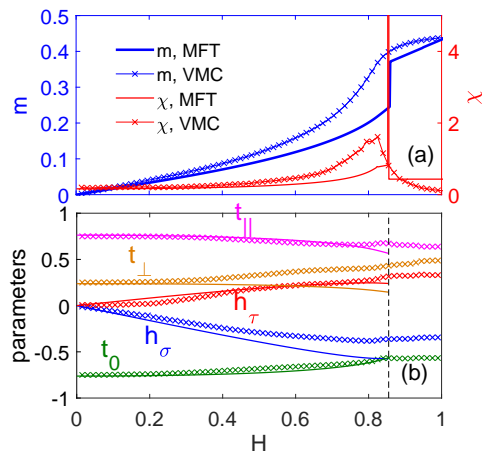} %{.png}
 \caption{(a) The induced magnetization $m$ (left scale) and susceptibility $\chi$ (right scale) versus $H$. The solid lines are from MFT and the symbols from VMC. (b) The MF parameters (solid lines) and VMC-optimized parameters (symbols) versus $H$. The vertical dashed line denotes $H_{c,MF}$, above which $t_{0,||,\perp}$ collapse to zero (while $h_{\si,\tau}$ are finite) in MFT.}\label{fig:systematics}
\end{figure}

We now discuss the properties of the ground state. The induced spin magnetization along the unit vector $(1,1,1)/\sqrt{3}$, $m=\sum_{\v r a} m_{a}(\v r)/(2N\sqrt{3})$, versus $H$ is shown in Fig.\ref{fig:systematics}(a). The MF result (blue solid line, left scale) develops a kink at $H_{c,MF}$, and accordingly a spike in the susceptibility $\chi=dm/dH$ (red solid line, right scale). In contrast, the magnetization from VMC (blue symbols, left scale) changes smoothly, and the susceptibility (red symbols, right scale) only shows a hump near $H_{c,MF}$. Therfore, from the magnetization point of view, the MF transition becomes a crossover in VMC. 
Microscopically, these behaviors are determined by the parameters in the effective Hamiltonians. Fig.\ref{fig:systematics}(b) shows the MF parameters (solid lines) and VMC-optimized parameters (symbols). They agree very well at low fields, but behaves differently at larger fields: In the MF case, the parameters $t_{0,||,\perp}$ all collapse to zero at $H>H_{c,MF}$ (where $h_{\si,\tau}$ begin to depend on site and direction), while all of the five parameters change smoothly in VMC. 

From perturbative analysis at low fields, it is known that the spin liquid has the Ising-topological order with Chern number $C=1$.~\cite{Kitaev2006, Nayak2011}. It is then interesting to ask whether this topology persists all the way to high fields, given the smooth change of the VMC parameters. To answer this question, we substitute the VMC-optimized parameters $X$ into \Eq{eq:cmft} to obtain $\cH_v = \frac{1}{2}\sum_\v k\psi_\v k ^\dag M_\v k\psi_\v k$ in momentum space. Since we find the energy bands of $M_\v k$ are nondegenerate except at isolated magnetic fields with gap closing, we can calculate the Chern number of each band, and we use the total Chern number $C$ of the valence bands of $M_\v k$ to define the overall topology. The Chern number is shown in Fig.\ref{fig:gap} (color bars), together with the
spinon excitation gap $\Del$ (symbols) obtained from the eigenvalues of $M_\v k$ over the Brillouin zone.
At low fields, the Chern number is $C=1$, and we checked $\Del$ scales as $H^3$, exactly as predicted by analytic perturbation theory.\cite{Kitaev2006} However, in the region $H_{c1}<H<H_{c2}$, where $H_{c1}\sim 0.83J$ and $H_{c2}\sim 0.92J$, the Chern number becomes $C=4$. Correspondinly, $\Del$ closes and reopens at the two magnetic fields  where the Chern number switches. Note the exact gap closing fields may lie between the discrete VMC data in Fig.\ref{fig:gap}, but they must be located where the Chen number changes. On the other hand, the gap $\Del$ from $M_\v k$ does not have an absolute meaning, since multipling $M_\v k$ by any positive factor does not change the trial ground state $|G\>$ for $|G_s\>$. However, we find the gap matches the MF result at low fields if we use the MF parameters as the input for VMC. 

Interestingly, the VMC Chern number returns to $C=1$ above $H_{c2}$, and becomes zero only when $H>H_{c3}\sim 1.83J$. Even at the latter stage, we find (not shown) that the variational parameters, and $t_{0,||,\perp}$ in particular, remain nonzero to gain energy. Therefore, while the system may be topologically trivial at high fields, spinon hopping and pairing on bonds still exist, in contrast to the case in the local moment state obtained from MFT.

Finally, despite of the departure at higher fields, our Z$_2$ MFT agrees with VMC exercellently up to, say, $H_{c,MF}/2$, a point we try to understand. Since the amplitude fluctuations of $K^a_{\mu\nu}(\v r,\v r')$ in \Eq{eq:mmft} should be gapped from a Landau theory point of view, we consider the sign changes, or in the jargon of gauge theory we attach a Z$_2$ gauge field to $K^a_{\mu\nu}(\v r,\v r')$. But it can be shown straightforwardly that if $K^a_{\mu\nu}(\v r,\v r')$ is a self-consistent MF solution, so is it after a local Z$_2$ gauge transformation, $K^a_{\mu\nu}(\v r,\v r')\ra s(\v r)K^a_{\mu\nu}(\v r,\v r')s(\v r')$. In fact they describe the same state of physical spins. This defines the Z$_2$ gauge structure of the MFT.\cite{Wen2004} So we can limit ourselves to those distinct sign changes of $K^a_{\mu\nu}$ that can not be removed by local Z$_2$ gauge transformations. This is the $Z_2$ vortex excitation. Numerics shows that the Z$_2$ vortex excitation is also gapped at finite fields in MFT.\cite{SM} (The fermions may or may not be gapped.) This suggests the robustness of the MF solution, up to gauge equivalence. In a more general context, our approach to the Z$_2$ spin liquid should be compared to the more sophisticated approach starting from a charge-SU(2) invariant Hamiltonian and the related SU(2) MF construction. If we used the symmetrized form for the spin operator, $\hat{\si}_a=i [c_a c_0 - \sum_{bc} \eps_{abc} c_b c_c/2]/2=\psi^\dag \si_a\psi/2$, in Eq.(\ref{eq:kitaev}), we would have $[Q_a,\cH]=0$, hence would gain explicit charge-SU(2) invariance. However, our practice shows that the corresponding SU(2) MF solution always conveges to the confined local-moment state (even at zero field), {\em if the local moment is included in the MF order parameters}. Therefore, starting from a high symmetry is neither a necessity nor an advantage. This adds value to the Z$_2$ MFT as a more efficient approach to construct a Z$_2$ spin liquid.

{\em Concluding remarks}: We investigated the spin liquid in the anti-ferromagnetic Kitaev model under a conical magnetic field ${\bf H}= (1,1,1) H$. From the VMC optimized effective Hamiltonian on the basis of a $\mathrm{Z_2}$ MFT, we found topological quantum phase transitions, with the Chern number changing from  $C=1$ to $C=4$, $C=1$ and $C=0$, successively with increasing field. The agreement between the Z$_2$ MFT and VMC may be understood by projective symmetry group arguments, and we proposed that the Z$_2$ MFT is an efficient approach to construct a stable Z$_2$ spin liquid. 
	
The spin-liquid states with different Chern numbers $C$ realize different topological orders, as summarized in Kitaev's 16-fold way~\cite{Kitaev2006}: The $C=1$ phase belongs to the (chiral) Ising topological order: It has chiral central charge $c=\frac12$ and anyon excitations in the superselection sectors $\{1,\si,\veps\}$, where $\si$ is a non-Abelian anyon with quantum dimension $\sqrt2$ and topological spin $1/16$, and $\veps$ is Abelian and has topological spin $1/2$ (namely it is a fermion). The $C=4$ phase has chiral central charge $c=2$ and Abelian anyon excitations in the superselection sectors $\{1,s_1,s_2,\veps\}$, where $s_1$ and $s_2$ are two chiral semions with trivial mutual statistics, and they fuse into the fermion $s_1\times s_2=\veps$. Therefore, the $C=4$ phase is a new type of spin liquids induced by an intermediate field, as compared to the $C=1$ phase known proviously from perturbation theory at small fields. The new spin liquid phase can be identified by measuring the modular matrices~\cite{WenSTMat} and the chiral central charge $c=C/2$ in future numerical studies, and by measuring the quantized thermal Hall conductivity~\cite{Matsuda2018} $\kappa_{xy}/T=c$ (in units of $\frac{\pi}6\frac{k_B^2}\hbar$) in future experimental studies.

\acknowledgements{MHJ and SL contributed equally to this project. QHW thanks Fan Yang for discussion on fast Pfaffian updates. The project was supported by the National Key Research and Development Program of China (under grant Nos. 2016YFA0300401 and 2015CB921700) and the National Natural Science Foundation of China (under Grant Nos.11574134, 11874115, and 11774152).}

\end{document}